\documentclass[12pt]{article}
\newif\ifams\amsfalse                                                    %%%
\amstrue                                                                 %%%
\ifams                                                                   %%%
 \usepackage{amsfonts}                                                   %%%
\fi                                                                      %%%
\newif\iffigs\figsfalse                                                  %%%
\figstrue                                                                %%%
\newif\ifdraft\draftfalse
\newif\ifinter\interfalse                                               
\intertrue                                                             
\ifdraft\else\interfalse\fi
\ifinter
 \else
\fi
\title{}
\ifinter
  \def\secl#1{\nopagebreak\marginpar{\vspace{-6mm}\scriptsize #1}\label{#1}}
  \def\beql#1{\marginpar{\vspace{4mm}\scriptsize #1}
              \nopagebreak\begin{equation}\label{#1}}
  \def\ftl#1#2{\footnote{\label{#1}[#1] #2}}
  \def\capl#1#2{\caption{[#1] #2}\label{#1}}
  \def\bibl#1{\marginpar{\vspace{4mm}\scriptsize #1}\nopagebreak\bibitem{#1}}
 \else
  \def\secl#1{\label{#1}}
  \def\beql#1{\begin{equation}\label{#1}}
  \def\ftl#1#2{\footnote{\label{#1}#2}}
  \def\capl#1#2{\caption{#2}\label{#1}}
  \def\bibl#1{\bibitem{#1}}
\fi

\def\draftnote#1%
  {\ifdraft
    \hspace{1mm}\newline\noindent\begin{tabular}[t]{|p{14cm}|}
     \hline {\bf DRAFT NOTE}\\ #1 \\ \hline
    \end{tabular}\newline\noindent
   \fi}

\def\internote#1%
  {\ifdraft\ifinter
    \hspace{1mm}\newline\noindent\begin{tabular}[t]{|p{14cm}|}
     \hline {\bf Internal Note}\\ #1 \\ \hline
    \end{tabular}\newline\noindent
   \fi\fi}

\def\multdn{$\downarrow\downarrow\downarrow\downarrow\downarrow$}
\def\beginsup%
 {\noindent\begin{tabular}[t]{|c|}
   \hline {\bf Supplements}\\ 
   \multdn\multdn\multdn\multdn\multdn\multdn
   \multdn\multdn\multdn\multdn\multdn\multdn \\ \hline
  \end{tabular}}

\def\multup{$\uparrow\uparrow\uparrow\uparrow\uparrow$}
\def\endsup%
 {\noindent\begin{tabular}[t]{|c|}
   \hline \multup\multup\multup\multup\multup\multup
          \multup\multup\multup\multup\multup\multup \\
   {\bf Supplements}\\ \hline
  \end{tabular}} 

\iffigs
 \input epsf
 \def\pct#1{\centerline{ \epsfbox{#1.eps}}}
\else
 \def\pct#1{(see figure in file #1.eps)}
\fi
\newif\ifappend\appendfalse
\appendtrue
\ifappend
 \newcommand{\newsection}[1]{
  \vspace{10mm} \pagebreak[3]
  \refstepcounter{section}
  \setcounter{equation}{0}
  \message{(\thesection. #1)}
  \addcontentsline{toc}{section}{\protect\numberline{\arabic{section}}{#1}}
  \begin{flushleft}
   {\large\bf\boldmath \thesection. #1}
  \end{flushleft}
  \nopagebreak}

\else
 \newcommand{\newsection}[1]{\section{#1}}
\fi

\def\al{\alpha}

\def\gm{\gamma}                
                \def\Dl{\Delta}

\def\lm{\lambda}               \def\Lm{\Lambda}
\def\om{\omega}               \def\Om{\Omega}
               \def\Sg{\Sigma}
\def\zt{\zeta}

\def\Mc{\mbox{\protect$\cal M$}}
\def\Nc{\mbox{\protect$\cal N$}}

\ifams
 \def\bbl#1{{\mathbb #1}}
\else
 \def\bbl#1l{{\bf #1}}
\fi

\def\RR{\bbl{R}}
\def\ZZ{\bbl{Z}}
\def\tr{{\rm tr}}

\def\pt{\partial}
\def\goto{\rightarrow}

\def\wbar{\overline}
\def\what{\widehat}
\def\dg{^{\dagger}}
\def\inv{^{-1}}
\def\rec#1{{\raise 0.4ex \hbox{$\scriptstyle {\frac{1}{#1}}$}}}

\def\half{{\raise 0.4ex \hbox{$\scriptstyle {1 \over 2}$}}}
\def\hs{\hspace{2mm}}
\def\hsc{\hspace{2mm},\hspace{5mm}}
\def\nl{\protect\newline}
\def\nlb{\protect\newline $\bullet$ }
\def\ie{{\em i.e.}}
\def\eg{{\em e.g.}}
\def\beq{\begin{equation}}
\def\eeq{\end{equation}}

\def\NPB#1{Nucl. Phys. {\bf B#1}}
\def\PLB#1{Phys. Lett. {\bf B#1}}
\def\PRV#1{Phys. Rev. {\bf #1}}
\def\PRP#1{Phys. Rep. {\bf #1}}

\begin{document}

%%
%% TITLE (either maketitle or manual)
%\maketitle

%\pagestyle{empty}
%% alternatively, if preface is one page long:
\begin{titlepage}

\ifdraft 
  \fbox{\bf !!!!!!!!!!!!!! 
  \ifinter INTERNAL \fi
  DRAFT VERSION !!!!!!!!!!!!!!}\vspace{-1cm}
\fi

\begin{flushright}
RI-5-98\\
CERN-TH-98-184\\
hep-th/9805123\\[5mm]
\ifdraft
 %%% calculate time 
 \count255=\time 
 \divide\count255 by 60
 \xdef\hourmin{\number\count255}
 \multiply\count255 by-60
 \advance\count255 by\time
 \xdef\hourmin{\hourmin:\ifnum\count255<10 0\fi\the\count255}
 %%% print date and time
 \number\day/\number\month/\number\year\ \ \hourmin
 %%% alternatively
 %\today
\\[10mm]\fi
\end{flushright}

\begin{center}
\Large
{\bf Aspects of Confinement and Screening\\
  in M Theory}
\\[5mm]
\large Shmuel Elitzur$^*$, Oskar Pelc$^*$
\normalsize and
\large Eliezer Rabinovici$^{*\dagger}$
\normalsize
\\[5mm]
{\em $^*$Racah Institute of Physics, The Hebrew University\\
  Jerusalem, 91904, Israel}\\
and\\
{\em $\dg$ Theory Division, CERN\\
  CH-1211 Geneva 23, Switzerland}\\
E-mail: elitzur@vms.huji.ac.il, oskar@shum.cc.huji.ac.il,\\
 eliezer@vxcern.cern.ch
\\[10mm]
\end{center}
%\begin{center}\bf Abstract}\end{center}
%\begin{quote}
%\end{quote}
%% alternatively, in titlepage:
\begin{abstract}
Confinement and Screening are investigated in SUSY gauge theories,
realized by an M5 brane configuration, extending an approach applied
previously to $\Nc=1$ SYM theory, to other models. The electric flux
tubes are identified as M2 branes ending on the M5 branes and the 
conserved charge they carry is identified as a topological property. 
The group of charges carried by the flux tubes is calculated and the
results agree in all cases considered with the field theoretical 
expectations. In particular, whenever the dynamical matter is
expected to screen the confining force, this is reproduced correctly
in the M theory realization. 
\end{abstract}
%-----------------------------------------------------------------

\internote{Figure \ref{f-qrk}:
$\gm_f\leftrightarrow\gm_c$ in .fig file; correct in .eps file}

%\pagestyle{plain}
%\setcounter{page}{1}
%% alternatively, if preface is one page long:
\end{titlepage}

\ifdraft
 \pagestyle{myheadings}
 \markright{\fbox{\bf !!!!!!!!!!!!!!
 \ifinter INTERNAL \fi
 DRAFT VERSION !!!!!!!!!!!!!!}}
\fi

%\tableofcontents
\flushbottom

%%%%%%%%%%%%%%%%%%%%%%%%%%%%%%%%%%%%%%%%%%%%%%%%%%%%%%%%%%%%%%%%
%% BODY OF PAPER
%%  use \newsection instead of \section

\newsection{Introduction}

The advance in the ability to analyze strong coupling effects in supersymmetric
field theories \cite{IS9509} led to a verification of some qualitative
ideas about the behavior of gauge field theories%
\footnote{See, for example \cite{Bander,Shuryak} and references therein.}.
In particular, confinement in $\Nc=1$ supersymmetric QCD (SQCD) was shown
\cite{SW94} to be related, as expected, to
monopole condensation.
\internote{Can we see magnetic screening? Is there a magnetic charge?}
In a confining phase, the electric field is confined to
flux tubes -- strings -- carrying a definite energy density (string tension).
Electrically charged sources are connected by such strings and this
leads to a confining force between them, \ie, a force that does not vanish as
the distance between them is increased.

An explicit demonstration of such a situation, for $\Nc=1$ SYM theory
was suggested recently, in the framework of a realization of
gauge theories in M theory on $\RR^{9+1}\times S^1$,
which is dual to type IIA string theory.
The 4D gauge theory is realized as the dimensionally reduced low energy
effective field theory on an M5 brane wrapped on a Riemann surface. 
When all the characteristic distances,
including the radius of $S^1$, are large with respect to the
Planck length, the semi-classical approximation is used \cite{Witten9703}.
Strictly speaking, the effective gauge theory is identified when the
radius $R$ of $S^1$ is vanishingly small,
corresponding to perturbative type IIA string theory.
As explained in \cite{Witten9706},
a change in $R$ has, in general, an effect on the 
world-volume field theory, so the results that one obtains for large $R$
are for a theory which can be different from the gauge theory
one attempts to study. 
However, there are indications that these theories are in the same
universality class and, therefore, have the same qualitative properties.
A weaker expectation would be that qualitative confining features,
will be shared by these theories.
It is those features which we wish to study in this paper.
\internote{See the Appendix for more details}

A candidate for flux tubes in $\Nc=1$ SYM theory was suggested in
\cite{Witten9706}: an M2 brane with one spatial dimension extended in a
physical direction and the other in an internal direction,
extending between points on the M5 brane.
This is called {\em an MQCD string};
it was explored further in \cite{HSZ9707}.
Considering the realization of $\Nc=1$ SQCD -- 
Supersymmetric $SU(N)$ gauge theory with fundamental quarks -- the quark
states were also identified as M2 branes ending on the M5 brane,
these with their full extension in internal space,
and it was shown that when confinement is
expected, these quarks cannot exist in isolation and must be either grouped
in multiples of $N$, forming baryons, or connected to MQCD strings, forming
mesons. Moreover, for the $\Nc=2$ model, weakly broken to $\Nc=1$ by a mass for
the adjoint, the authors of \cite{HSZ9707} identified $N-1$ types of strings
with different tensions and reproduced the field theoretical results of 
\cite{DS9503}.

The string connecting two oppositely charged sources, carries flux which is
determined by the sources.
In the absence of dynamical matter,
this flux protects the string from breaking, by charge conservation.
However, when there is dynamical matter carrying an appropriate charge,
a pair may be created from the vacuum, cutting the string in two.
Physically, this means that the potentially confining force between the
external sources is screened by the dynamical matter.
Because of this screening, the string itself is expected to be charged only
under those elements of the gauge group that act {\em trivially} on
the dynamical fields. In particular, when there are no such group elements,
one does not expect stable strings.
Instead, the force between any external sources is
expected to vanish as the distance between them increases.

\internote{See the appendix for more about the distinction between phases.}
This leads to a distinction between two physically different possibilities,
depending on the algebraic structure of the matter sector.
When the dynamical matter can screen any external charge, \ie, when there
are no group elements acting trivially on the dynamical fields, the forces
felt by these charges are qualitatively the same as in the Higgs phase
and there is no actual phase boundary between the Higgs and confining phases;
the corresponding branches are smoothly connected \cite{Higgs-Conf}.
On the other hand, when there are external charges that cannot be screened
by the dynamical matter, the force between them is qualitatively different
in the Higgs and Confining branches and, therefore, these branches are
expected to be separated by a boundary.
To determine which of these possibilities is realized, one should find
the subgroup $H$ of the gauge group that acts trivially on the dynamical
fields. 
Confinement and Higgs phases are expected to be distinct iff $H$ is
non-trivial.
A realization of the subgroup $H$ was suggested in \cite{Witten9706}.
Considering $\Nc=1$ SYM theory,
$H$ was identified as the homology group of the MQCD string and
this group was, indeed, shown to be isomorphic to $H=\ZZ_N$.
\internote{In the $N=4$ duality, $Z(G)=\pi_1(\tilde{G})$. Is there a relation?}
In this work we elaborate on the geometric manifestation
of the kinematic algebraic screening considerations, implied by the above
identification of $H$: the confining phase should be distinct from the Higgs
phase, when the later exists,
iff there is a stable MQCD string, carrying a non-trivial $H$ charge.

To investigate confinement and screening, it is
useful to realize the external probes as dynamical, but very heavy,
particles, as is done, for example, in \cite{HSZ9707}.
At energies small compared to the mass of the probes
\ie, when the string connecting them is not too long%
\footnote{The distance between the probes must, however, be large enough,
to justify the string description.},
it is energetically protected from
being cut by a pair of the massive particles. 
Therefore, such a string probes the model without the heavy particles:
if in this model confinement and Higgs phases are distinct,
this will be manifested by the stability of the string.
When the string is long enough, it should become unstable.

In the present work we use M theory to study confinement and
screening for systems with various local and global symmetries.
We identify the MQCD string, find its conserved charge -- the homology group
-- and compare to the field-theoretical expectation -- the subgroup $H$.
We then introduce external quarks and
demonstrate confinement, when they are heavy and screening, when they are
light. All our results are in agreement with the field theoretical
analysis.
The outline of this work is as follows.
In section \ref{sec-SU} we consider $SU(N)$ models.
We start with $\Nc=1$ SQCD, the model
considered in \cite{Witten9706,HSZ9707}, as the simplest example, to explain
the relevant concepts and methods. We then add an adjoint matter field
with a polynomial superpotential.
Section \ref{sec-lSU} is devoted to models with products of $SU(N)$
gauge factors and section \ref{sec-O4},
to models with $SO(N)$ and $Sp(N)$ gauge groups.
The $SO$ and $Sp$ models are realized here by configurations that include
an orientifold 4-plane, and we find ourselves on an unpaved way as far as
brane dynamics is involved. Guided by field theoretical expectations, we
suggest some rules for M2 brane configurations in the presence of the
orientifold. %
\internote{expand (?):
\nl We conjecture that the M2 brane configuration must be symmetric under the
orientifold transformation.
Apart from being intuitively reasonable, we present evidence
for the validity of this conjecture
from perturbative string theory and from the quantum Coulomb
branch of the $\Nc=2$ gauge theory.}
These rules are then used to obtain predictions for the confining
behavior of all the $SO$ and $Sp$ models considered.
We end in section \ref{sec-disc} with a summary and concluding remarks.
In the appendix we prove some properties of M2 branes used in section
\ref{sec-O4}.

\newsection{The $SU(N)$ Models}
\secl{sec-SU}

In this section we consider models with $SU(N)$ gauge group. We start with
a brief review of the brane realization%
\footnote{For a recent review, see \cite{GK9802}.}
and then analyze the $\Nc=1$
SYM model. We use this model to introduce and explain the concepts and methods
that are used in the other parts of this work.
In subsection \ref{sec-screen}, we show that the dynamical fundamental quarks
screen the confining force and in subsection \ref{sec-adj},
we add an adjoint matter field with a polynomial superpotential.

\subsection{The Brane Configuration}
\secl{SU-brane}

We start with the brane configuration in weakly coupled type IIA string
theory. It is illustrated in figure
\ref{f-cfg}.
\begin{figure}
\pct{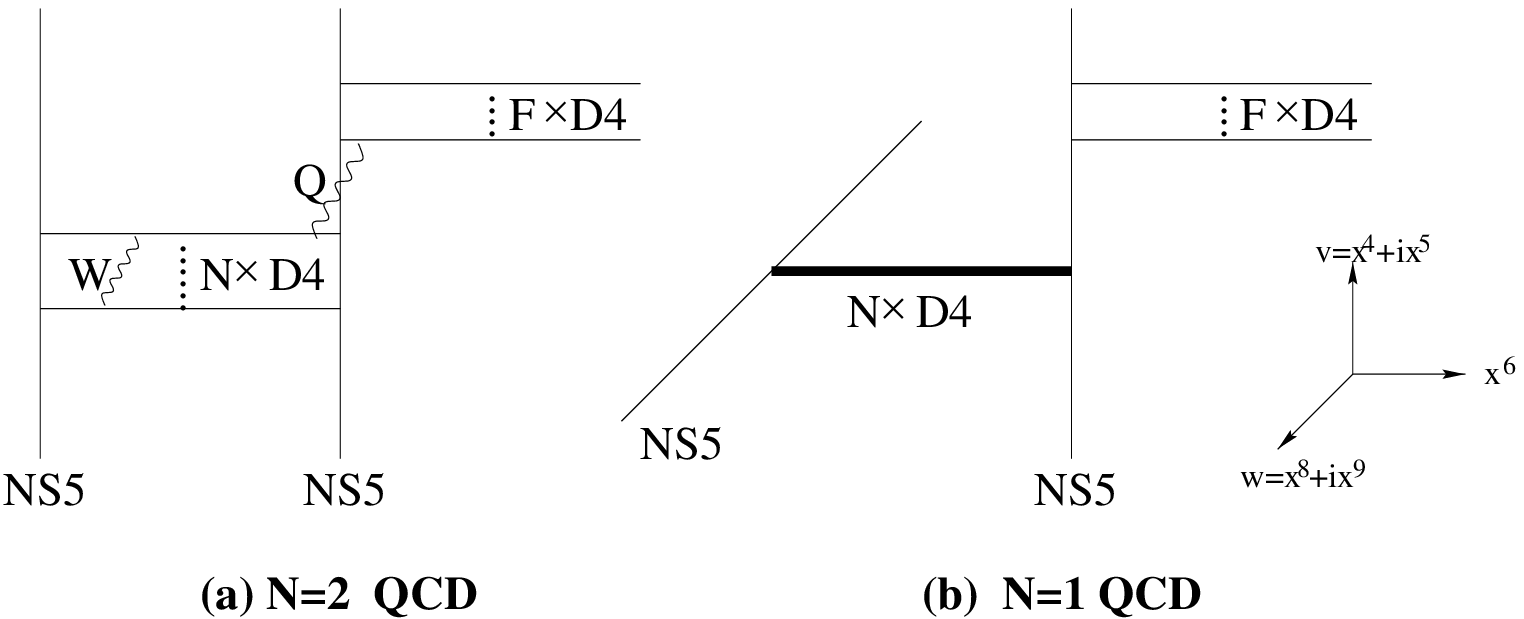}
\capl{f-cfg}{The SUSY $N=2$ and $N=1$ brane configurations}
\end{figure}
All branes are extended in the $x^0-x^3$ directions, and are at a fixed value
of $x^7$,
so these directions are suppressed. There are two NS5 branes at fixed values of
$x^6$ while the D4 branes are extended in this direction. $N$ of them are
suspended between the two NS5 branes and $F$ additional D4 branes end on the
right NS5 brane and extend%
\footnote{For a more complete description of the gauge theory, including the
Higgs branch, the semi-infinite D4 branes should actually end on D6 branes
extending in the $x^7,x^8,x^9$ directions. In the description used here,
these D6 branes are taken very far away in the $x^6$ direction.}
to $x^6\goto\infty$.

When the NS5 branes are parallel (\eg, extend in the $v=x^4+ix^5$ direction;
see figure \ref{f-cfg}a),
the low energy effective field theory on the world-volume
of the {\em finite} D4 branes, \ie, those with  
a finite extent in the $x^6$ direction, is an $\Nc=2$ 4D $SU(N)$ gauge theory
with $F$ hypermultiplets in the fundamental representation \cite{HW9611}.
The vector multiplets, in the adjoint representation,
originate from strings between the finite D4 branes,
while the hypermultiplets originate from strings between a
semi-infinite D4 brane and a finite one.
The finite D4 branes are free to move in the $v$ direction,
with the `center of mass' frozen by a non-perturbative effect
\cite{Witten9703},
and this motion realizes the Coulomb branch of the field theory:
the $v$ locations of these branes are identified with the eigenvalues of the 
adjoint scalar's vev.

A rotation of one of the NS5 branes in the $v,w$ space, where $w=x^8+ix^9$,
corresponds to a mass for the adjoint scalar, which lifts the Coulomb branch.
Indeed, such a rotation restricts the
finite D4 branes to the unique ($v,w$) point shared by both NS5 branes.
Generically, such a rotation will break all remaining
supersymmetry, but there are rotations that preserve half of the remaining
supersymmetry \cite{BDL9606}, corresponding to an $\Nc=1$ supersymmetric mass
term for the adjoint hypermultiplet \cite{Barbon9703}. 
In particular, when the right NS5 brane is extended in the $w$ direction
(see figure \ref{f-cfg}b),
the adjoint scalar and its spinor superpartner decouple -- get infinite
mass -- and one obtains \cite{EGK9702} $\Nc=1$ SQCD.

As already noted,
the above description is in the framework of weakly coupled type IIA string
theory, which is a limit of M theory compactified on $\RR^{9+1}\times S^1$,
with vanishingly small radius $R$ of the circle $S^1$.
When $R$ is finite,
the above brane configuration turns into
a single M5 brane \cite{Witten9703} and for large $R$, compared to the
11D Planck scale, one uses M theory.
The M5 brane is of the form $\RR^{3+1}\times\Sg$,
where $\RR^{3+1}$ corresponds to the $x^0-x^3$ directions and $\Sg$ is a two
dimensional surface embedded in the internal space $\Mc$, parameterized by the 
coordinates $v,w$ and $s=x^6+ix^{10}$.

In the $\Nc=2$ configuration \cite{Witten9703},
the NS5 branes contribute to $\Sg$ two surfaces in the $v$ direction and each
D4 brane contributes a tube, wrapping once the $x^{10}$ direction
(degenerating to a line in the $R\goto0$ limit).
The resulting curve is the Seiberg-Witten curve \cite{SW94}, encoding the
low energy effective gauge coupling. The various BPS states are realized as 
minimal membranes -- M2 branes -- with their full extension in $\Mc$,
\ie, point-like in $\RR^3$, and having boundaries on the M5 brane.
When the NS5 branes are not parallel, as in figure \ref{f-cfg}b,
the finite D4 branes merge and form a single
tube wrapping the compact dimension $N$ times.
For the configuration described in figure \ref{f-cfg}b,
when the NS5 branes are orthogonal to each other,
$\Sg$ assumes the following form \cite{HOO9706,Witten9706,BIKSY9706}:
\beql{Sg-SU}
t=v^{-N}\det(m-v)/\det m \hsc vw=\zt:=[\Lm^b\det m]^{\frac{1}{N}} \hs.
\eeq
Here $t=t_0e^{-s/R}$ ($t_0$ being a dimension-full constant), $\Lm^b$ is the
instanton factor, $b=3N-F$ and $m$ is the quark mass matrix. Geometrically,
the eigenvalues of $m$ are the $v$-locations of the semi-infinite D4 branes.
The curve (\ref{Sg-SU}) is only valid for $\det m\neq0$, but
we indeed consider only massive quarks.

\internote{UNITS: Hanany et al 9707 (Appendix); Oz et al 9711 (Appendix)}

\subsection{The MQCD string}
\secl{sec-hom}

In $\Nc=1$ SYM theory ($F=0$), all the dynamical fields are in
the adjoint representation.
The kernel $H$ of this representation (\ie. the set of
group elements acting trivially) is the center $\ZZ_N$, so one expects the
existence of strings (flux tubes) with a conserved $\ZZ_N$ charge. 
This means that a single string (with unit charge) should be stable
(protected by its charge), but $N$ parallel strings should be able to
annihilate.

A realization of such a string, called MQCD string, was studied in
\cite{Witten9706}.
\internote{Extended objects in MQCD were studied also in
[Volovich9710120,FS9711083].
Those were BPS objects. In contrast, the present one is not BPS, since it
does not carry an additively conserved central charge (which could appear in
the SUSY algebra).}
It is an M2 brane with its boundary on the M5 brane.
In its simplest form, its world-volume is of the form
$\RR^{1+1}\times C$, where the factor $\RR^{1+1}$ is a string
in the physical spacetime $\RR^{3+1}$ and $C$ is
a curve in the internal space $\Mc$, extending between points on $\Sg$.
Its $\ZZ_N$ charge was identified with the homological classification of
such surfaces.
In the following we recall this analysis, relying on a pictorial
approach, which will be useful in the later derivations.

The relevance of the homological classification to conserved charges can
be motivated as follows:
the homological equivalence of two brane configurations means that there
is a world-manifold that can be interpreted as describing an evolution
between them, so a property which is the same for equivalent configurations
will be conserved under such an evolutions.
This is illustrated in figure \ref{f-toy},
\begin{figure}
\pct{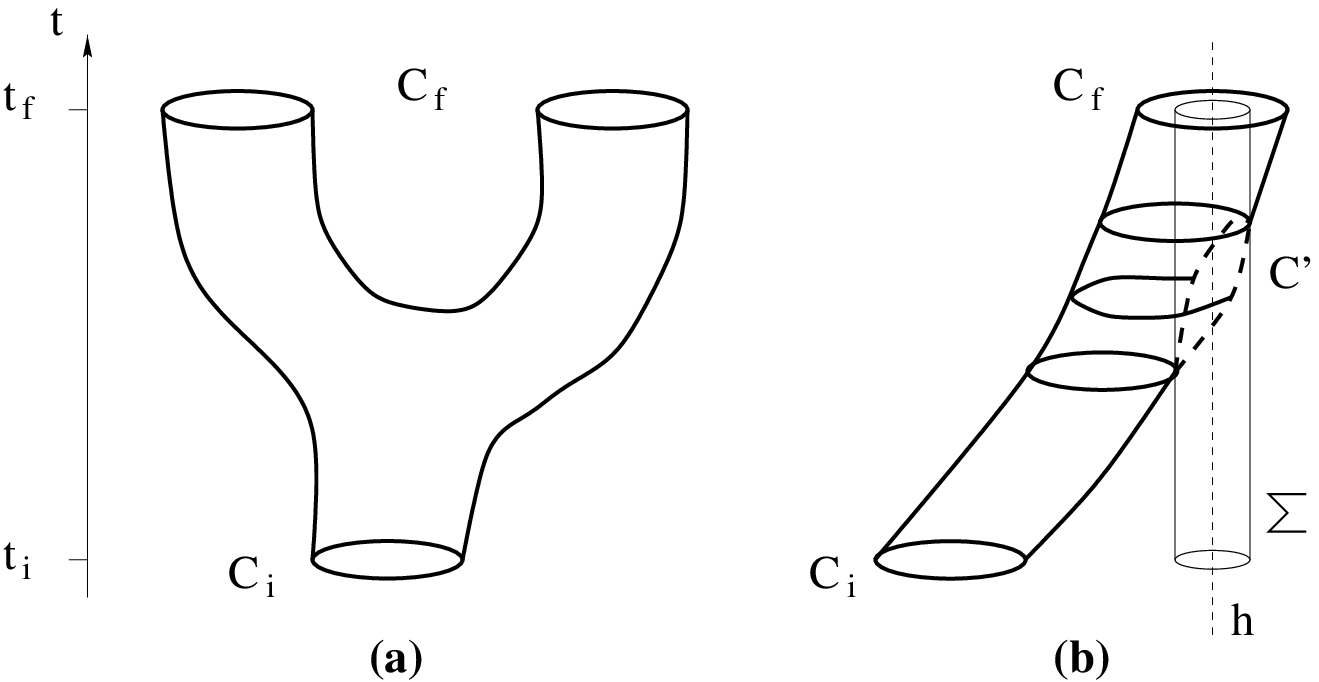}
\capl{f-toy}{The world-sheet of a string, evolving in a punctured plane
\nl (the vertical dashed line $h$ is the world-line of the `hole' in space).
\nl (a) a closed string: $C_i$ and $C_f$ have the same (vanishing) winding
number around the hole; 
\nl (b) a string that is allowed to end on a circular D1-brane $\Sg$: this
allows different winding numbers for $C_i$ and $C_f$.}
\end{figure}
describing the evolution of a string in $2+1$ dimensions. 

When one considers a closed $p$-brane (with no boundary), the above
reasoning leads to the usual homology group%
\ftl{time}{Strictly speaking, one is lead to $H_p(\Mc\times\RR)$, where
$\RR$ denotes the time axis, but $\RR$ is trivial topologically.}
$H_p(\Mc)$: {\em two $p$-branes%
\footnote{We consider here homology groups with {\em integer} coefficients;
an element of such a group is a {\em sum} of manifolds which can always be
viewed as a `single $p$-brane', possibly disconnected and/or `wrapped' a
few times.} 
$C_i$, $C_f$ (in the space $\Mc$) are equivalent if the difference
$C_f-C_i$ is a boundary of a $(p+1)$-brane.} 
For example, figure \ref{f-toy}(a) describes the evolution of a closed
string from the initial configuration $C_i$ (consisting of a single string)
to the final one $C_f$ (consisting of two strings) and $C_i$ and $C_f$ are
homologically equivalent.
To see what this equivalence means, assume that the space
$\Mc$ in which the string evolves is a punctured plane:
the line $h$ in figure \ref{f-toy} describes the world-line of the
excluded point.
A closed string has a conserved winding
number around this point (which is an element of $H_1(\Mc)=\ZZ$) and $C_i$
and $C_f$ have {\em the same} winding number \ie, the winding number is
conserved.

When the $p$ brane is allowed to end on some manifold $\Sg$, the boundary of
its world-manifold $S$ can be more complicated, as illustrated in
figure \ref{f-toy}b: in addition to the initial and
final configurations ($C_i$ and $C_f$) there may be another part $C'$ which
describes the evolution of the $p$-brane's boundary. This part is constrained
to be on%
\footnote{More precisely, $C'$ is restricted to the {\em world-manifold}
of $\Sg$, however, assuming $\Sg$ is static, this world-manifold is
$\Sg\times\RR$ and the $\RR$ factor is trivial topologically.} 
$\Sg$, so one is led to the {\em relative homology group}, denoted by
$H_p(\Mc/\Sg)$:
{\em two $p$-branes $C_i$, $C_f$ are equivalent if the difference
$C_f-C_i$ is either a boundary or can be complemented by another
$p$-brane $C'$, which lies in $\Sg$, to form a boundary.}
The significance of this change in classification is demonstrated in figure
\ref{f-toy}b: the string is allowed to end on a circular D1-brane $\Sg$,
which surrounds the excluded point.
As a result, the winding number is no longer conserved
($C_i$ and $C_f$ in figure \ref{f-toy}b have {\em different} winding number).
This is expressed mathematically by the fact that the relative homology group
$H_1(\Mc/\Sg)$ is trivial in this case. One can see clearly that this lack of
conservation is closely related to the fact that $\Sg$ itself carries a
non-trivial winding number.

We return now to our specific case: the MQCD string. It is an M2 brane which
looks in the physical space as a string, extending between external charges at
infinity. In its simplest form, it has the structure
$\RR\times C$, where the factor $\RR$ represents the coordinate $x$ along the
string and $C$ is a curve in the internal space $\Mc$. More generally, $C=C_x$
can depend on $x$.
We look for a conserved property of this string that can
be identified with the flux of the flux tube that this string represents, so
(1) this should be a property of the {\em cross-section} $C_x$ of the string,
which is independent of $x$ and (2) it should governs the stability of the
string.
The homological classification of the cross-section obeys both these
requirements: (1) different cross-section are, by definition, homologically
equivalent and (2) to break the string, the cross-section must be deformed to
a point, which is possible iff the cross-section is trivial homologically.
Therefore, one is led to identify the charge group $H=\ZZ_N$ of the flux tube
with the homology group $H_1(\Mc/\Sg)$.
  
To show (following \cite{Witten9706}) that $H_1(\Mc/\Sg)$
is indeed isomorphic to $\ZZ_N$, 
we first state a general result concerning $H_1(\Mc/\Sg)$,
which is true whenever $\Sg$ is connected
(as will be the case in all the models we will consider).
\begin{itemize}
\item
To Investigate $H_1(\Mc/\Sg)$, it is enough to consider {\em closed curves},
since an open curve $C$ (ending on $\Sg$) can be complemented to a closed one
by a curve in $\Sg$ connecting the endpoints of $C$.
\item
To obtain $H_1(\Mc/\Sg)$, one can start from $H_1(\Mc)$ and find the elements
(equivalence classes) that contain (closed) curves in $\Sg$. These elements
form a subgroup $K$ of $H_1(\Mc)$. The curves in $K$ are exactly the curves
which are trivial in $H_1(\Mc/\Sg)$, therefore,
\beql{HR}  H_1(\Mc/\Sg)\simeq H_1(\Mc)/K \hs. \eeq
\end{itemize}
This result can be derived \cite{Witten9706} using the exact sequence
\[ H_1(\Sg)\goto H_1(\Mc)\goto H_1(\Mc/\Sg)\goto H_0(\Sg). \]

Now we turn to the specific form of $\Mc$ and $\Sg$ and apply eq. (\ref{HR}).
Recall that $\Mc$ is of the form $\RR^5\times S^1$, where $S^1$ is the 
10th (compact) spatial dimension. $\Mc$ is parameterized by $v,w$, and
$t=e^{-(x^6+ix^{10})/R}$, so arg($t$) parameterizes the compact dimension
and the point $t=0$ (corresponding to $x^6=\infty$) is excluded.
Therefore, $H_1(\Mc)=\ZZ$,
its elements being the winding numbers around $t=0$.
The embedding of $\Sg$ in $\Mc$ is given by
\beql{Sg-SU0} t=v^{-N} \hsc vw=\zt \hs, \eeq
so $\Sg$ can be identified with the punctured $v$ plane (with the origin
deleted).
Therefore, closed curves in $\Sg$ are classified by their winding number
around $v=0$ and from the relation $t=v^{-N}$ in eq. (\ref{Sg-SU0}) we learn
that each loop around $v=0$ is accompanied by $N$ loops around $t=0$. 
This means that the group $K$ of elements in $H_1(\Mc)$, which are trivial in
$H_1(\Mc/\Sg)$ are the multiples of $N$ so, using eq. (\ref{HR}), we obtain
$H_1(\Mc/\Sg)=\ZZ_N$.

To summarize, the MQCD string is an M2 brane which is represented in the
internal space $\Mc$ by a curve $C$, extending between points in $\Sg$.
As such, it has a (topological) $\ZZ_N$ charge,
that is identified with the center of the gauge group.
To determine the charge of a given string $C$ one should
`close' it by a curve in $\Sg$ and find the winding number of the resulting
curve around $t=0$.

Such a string is described in figure \ref{f-sus}.
\begin{figure}
\pct{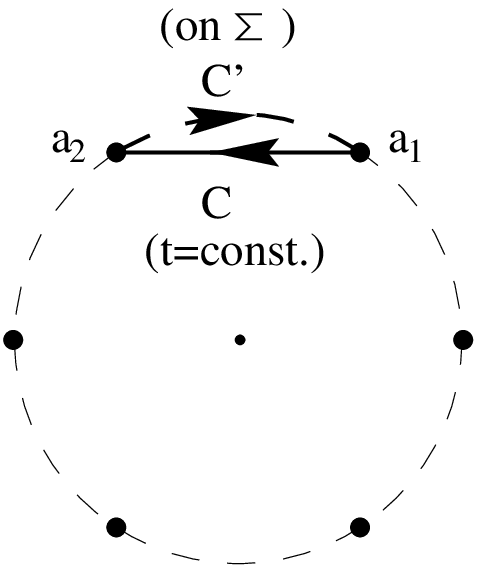}
\capl{f-sus}{The MQCD string}
\end{figure}
This is a projection on the $v$ plane. Each point in the figure corresponds
to a unique point on $\Sg$ but we want to consider also points in
$\Mc$ which are {\em not} on $\Sg$.
To distinguish between them we use dashed lines
for curves on $\Sg$ and full lines for curves not on $\Sg$. The circle
is on $\Sg$ and corresponds to $|v|=$const., which implies $x^6=$const.. The
emphasized points have also common $x^{10}$ and there are $N$ such points.
Now we connect two adjacent points $a_1,a_2$ by a curve $C$ with
{\em constant $t$} (which is, therefore, not in $\Sg$).
To find its charge, we close it by adding to it the arc $C'$ above it (which
is in $\Sg$). The resulting loop winds once around $x^{10}$, so $C$ has unit
charge. Note that we could close
$C$ from below, obtaining the charge $-N+1$, so the charge is indeed defined
only modulo $N$.

The loop $C+C'$ can be moved off $\Sg$ (having no boundary),
and then there is no obstruction from shrinking it to a point in the $v,w$
direction. However, it cannot disappear, since it still winds around $x^{10}$.
Instead, we have obtained a cylindrical M2 brane, winding around $x^{10}$
and such an M2 brane is identified as the fundamental closed string. 
This is a good place to make an important remark.
The process we have described demonstrates that the MQCD string is equivalent
{\em topologically} to a closed string.
This means that there is a {\em kinematical} possibility that one will
transform to the other, however,  such a process may be highly suppressed
by the dynamics.
In particular, as explained in \cite{Witten9706},
to be able to investigate the MQCD string while ignoring stringy effects,
we should be in a range of parameters where an MQCD string
{\em cannot} transform to a closed string. This range of parameters was
explored in \cite{Witten9706}.
In this work we are mainly concerned with kinematics, as manifested by the
topology of world-manifolds. So when we say that two curves are equivalent,
it should be understood in this kinematical sense.

\subsection{Confinement}
\secl{sec-conf}

As explained previously, the charge of the MQCD string protects
it from breaking, so in the absence of $\ZZ_N$ charged particles (quarks),
it is infinite in physical space $R^3$ and is stable (infinite in time).
This is interpreted as a signal of confinement in the gauge theory.
To investigate confinement further, one introduces heavy fundamental matter
(quarks) that, at this stage, is used to provide external probes of the
gauge force \cite{HSZ9707}.
This is done, as described in the beginning of this section,
by adding semi-infinite D4 branes -- tubes of M5 brane in M theory.
The large mass means that the D4 branes
are very far (in the $v$ direction) from the finite D4 branes. This implies
that they have negligible influence on the geometry of $\Sg$ at this region
and the analysis of the previous subsection continues to apply. In particular,
the MQCD string still exists and, at low enough energies (at which it is
restricted to be kept far from the semi-infinite D4 branes) it has the same
stability properties. 

To identify the quarks, one considers the $\Nc=2$ model in the Coulomb phase,
where the unbroken gauge symmetry is Abelian and the quarks are not
confined.
The corresponding BPS states are identified with M2 branes with
a boundary that surrounds (in the $v$ plane) one finite and one semi-infinite
D4 brane%
\footnote{We assume here that all the masses are distinct. To realize all the
quark states when masses coincide, one has to introduce D6 branes.};
in principle, it may be either a disk (with one boundary) or a cylinder
(with two boundaries, each winding one D4 brane)%
\footnote{M2 branes representing BPS states were discussed in
\cite{FS9706,HSZ9707,HY9707,Mikhailov9708,NOY9712}. In particular, 
their topology was considered in \cite{HY9707,Mikhailov9708}.}.
\internote{This may require some clarification. In the absence of M5 branes,
a fundamental string is represented in M theory by a cylindrical M2 brane
which wraps once around $x^{10}$, so one would expect that the quark
M2 brane would be cylindrical. However, when the M2 brane ends on an M5
brane, in certain situations it can deform `continuously' (in the homological
sense) between a disk and a cylindrical topology (such a transition was
described, in the previous subsection, where the MQCD string was transformed
to a fundamental string).
This is the situation with the membrane representing a quark.
There is evidence \cite{HY9707} that in certain situations the (minimal) M2
brane corresponding to a quark is a disk,
however, moving in parameter-moduli space, one can obtain situation where
such a disk must deform to a cylinder.}
These two types of surfaces are illustrated in figure \ref{f-qrk}.
\begin{figure}
\pct{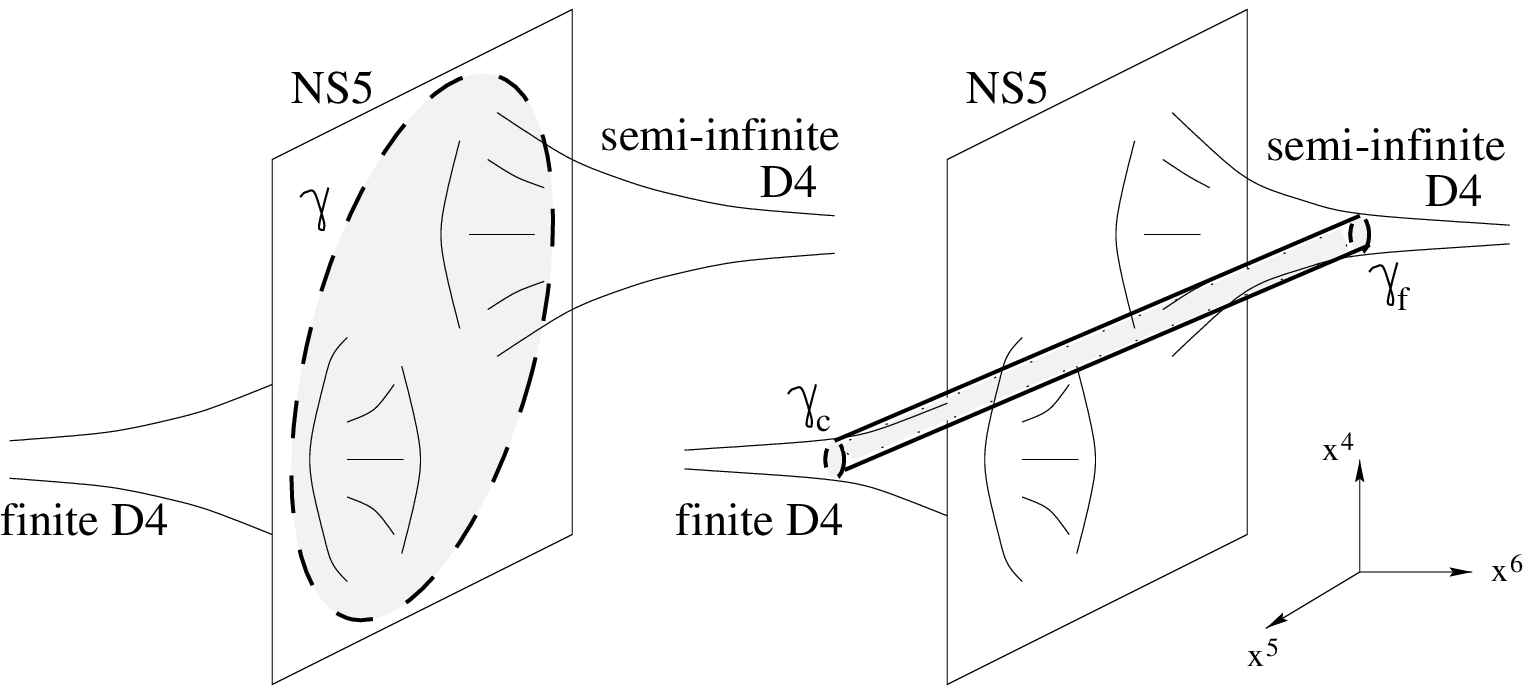}
\capl{f-qrk}{M2 branes representing quarks. 
\nl The thin lines describe an M5 brane, the thick lines -- an M2 brane
and the dashed lines, boundaries of the M2 brane on the M5 brane.
\nl The left figure describes a surface with a disk topology and the
right one -- a cylindrical topology.}
\end{figure}
They are  `continuously' connected, \ie, they are homologically equivalent.
In the following, we will refer only to the cylindrical topology, but we do
not loose generality by this, since one can deform `continuously' (in the
above homological sense) between the two topologies.
In particular, the unique boundary loop $\gm$ of a
disk can be seen as a limiting case of a two-loop boundary $\gm_c+\gm_f$ of a
cylinder. 
The winding numbers of the boundary (in the $v$ plane) around the D4 branes
represent the color and flavor quantum numbers of the quark (the two possible
orientations correspond to a quark and anti-quark).

Next one rotates the left NS5 brane. In field theory, this corresponds to a
mass for the adjoint hypermultiplet. As a result, monopoles condense and this
is expected to trigger confinement of electric charge.
This indeed can be seen in the brane configuration \cite{HSZ9707}:
Upon rotation, all the $N$ finite tubes (D4 branes) join to a
single tube, winding $N$ times around $x^{10}$. As a result, the M2 brane
configurations described in figure \ref{f-qrk} are not possible any more.
To understand this, note that $\gm_f$ winds once around $x^{10}$, so $\gm_c$
should also wind once, but there is no such loop (with color quantum numbers).
This is seen as a manifestation of confinement: there are no isolated
quarks. To see what {\em is} possible, one may check what happens to $\gm_c$,
as the left NS5 brane is rotated.
This is described (for $SU(2)$) in figure \ref{f-cnf}.
\begin{figure}
\pct{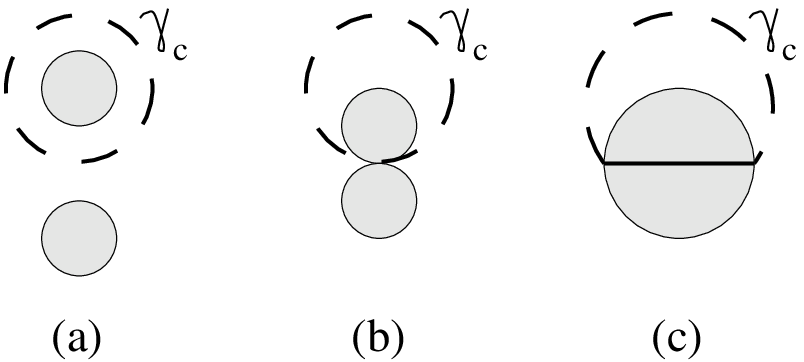}
\capl{f-cnf}{The fate of the quark M2 brane.
\nl (a) the $\Nc=2$ Coulomb phase,
\nl (b) on the verge of confinement,
\nl (c) the $\Nc=1$ model: the continuous line is {\em not} on $\Sg$.}
\end{figure}
This is a schematic%
\footnote{It is {\em schematic}, since it mixes $v$ and $x^{10}$ directions.
More details can be found in \cite{HSZ9707}.}
illustration of a cross-section of the M5 brane between the two NS5 branes.
The shaded circles represent the
D4 brane tubes. As the tubes join, $\gm_c$ is forced to have a part that
{\em is not} on $\Sg$. Since the M2 brane cannot end there, it must connect
to another piece of an M2 brane. Comparing with figure \ref{f-sus},
one can see that this other piece can be the MQCD string with a unit charge.
Indeed, the curve $C+C'$ in figure \ref{f-sus} winds once around $x^{10}$,
exactly as is required from $\gm_c$. To summarize, we have the following
evidence for confinement in $\Nc=1$ MQCD:
\begin{itemize}
\item in the absence of external quarks, the MQCD string is infinite and
stable;
\item M2 branes representing isolated quarks do not exist;
\item there is an M2 brane representing a meson: quark and anti-quark with an
MQCD string extending between them.
\end{itemize}
\internote{Related information:
\nlb The non-existence of isolated quarks  can be traced to the condensation
of monopoles.
\nlb M2 branes representing a baryon can also be constructed
\cite{HSZ9707}.}

\subsection{Screening}
\secl{sec-screen}

To investigate screening, we consider higher energy scales, which means that
we allow more energetic excitations and, in particular, a longer string.
In the M theoretic description it means that we allow M2 branes with
a larger area, so the region of the semi-infinite D4 branes
becomes accessible to the M2 brane.
Therefore, the stability of the MQCD string should be
re-examined now, with the full curve (\ref{Sg-SU}) replacing 
(\ref{Sg-SU0}) as $\Sg$.
To use relation (\ref{HR}), all we have to do is to find
which closed curves in $\Mc$ are equivalent
(in $H_1(\Mc)$) to curves in $\Sg$.
$\Mc$ is unchanged by adding quarks (only the M5 brane is modified),
thus $H_1(\Mc)$ is still generated by a loop around the compact direction.
The semi-infinite D4 brane which represents a quark winds
{\em once} around the compact dimension, as can be easily verified%
\footnote{Recall that we consider distinct masses. With coinciding masses,
we would introduce D6 branes, which would make $\Mc$ simply connected,
leading to the same result for $H_1(\Mc/\Sg)$.},
therefore, {\em all} closed curves are equivalent to curves in $\Sg$,
so $H_1(\Mc/\Sg)$ is the trivial group. This means that in this case $\Sg$
leaves no residual conserved charge from the winding number in $\Mc$
and the MQCD string is not protected anymore by topological considerations
-- it can break. Moreover, it is clear that the semi-infinite D4 brane
should be involved in this process. 

In the following we demonstrate how the MQCD string can be broken.
We consider a world-manifold that will describe such a process or,
equivalently, a family of M2 branes parameterized by time. This is described
in figure \ref{f-scr}.  
\begin{figure}
\vspace{-1cm}
\pct{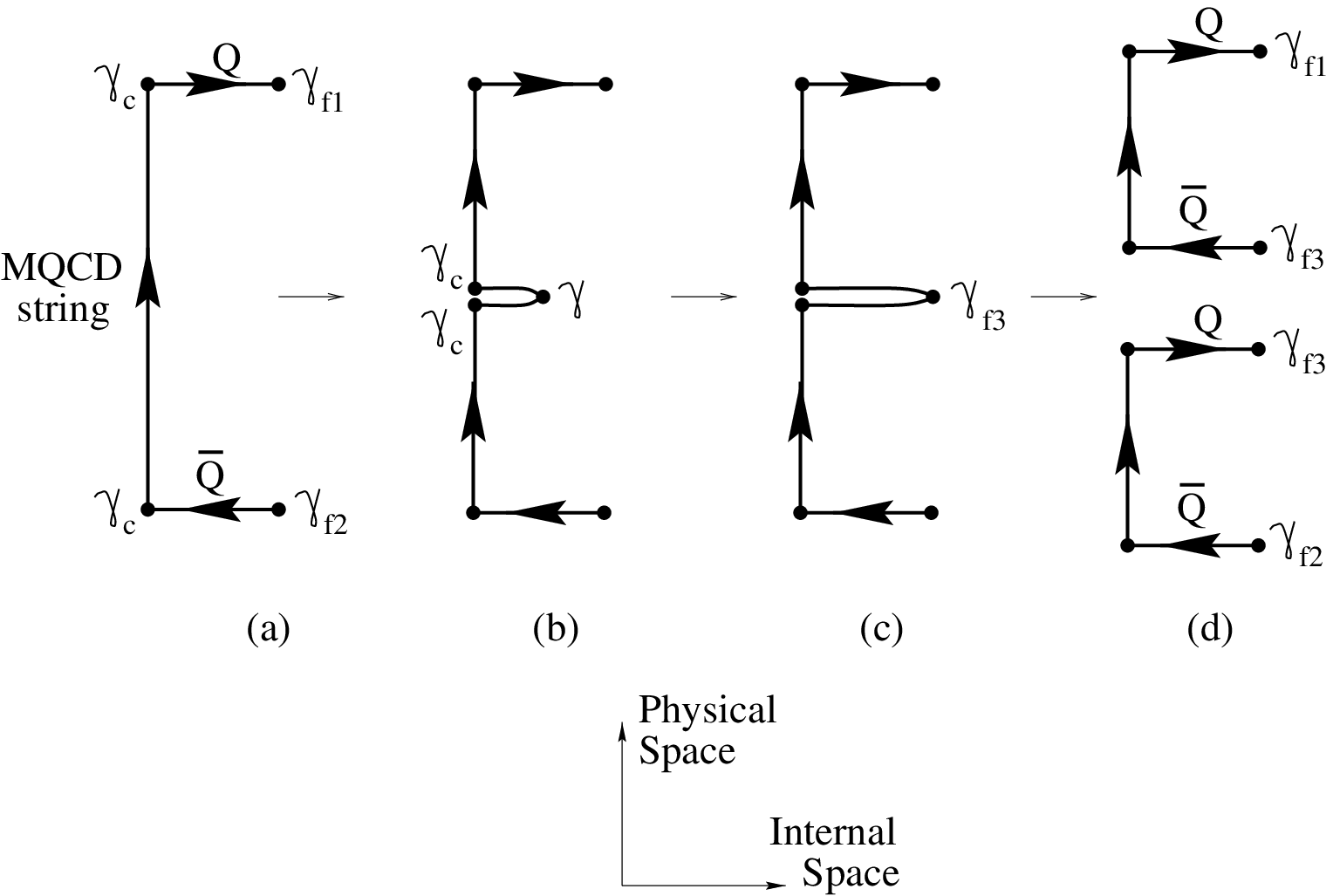}
\capl{f-scr}{Screening: breaking the MQCD string by a $Q\bar{Q}$ pair 
creation.
\nl (a) a meson: quark and anti-quark connected by an MQCD string;
\nl (b) an intermediate situation;
\nl (c) the final situation;
\nl (d) two disconnected mesons.}
\pct{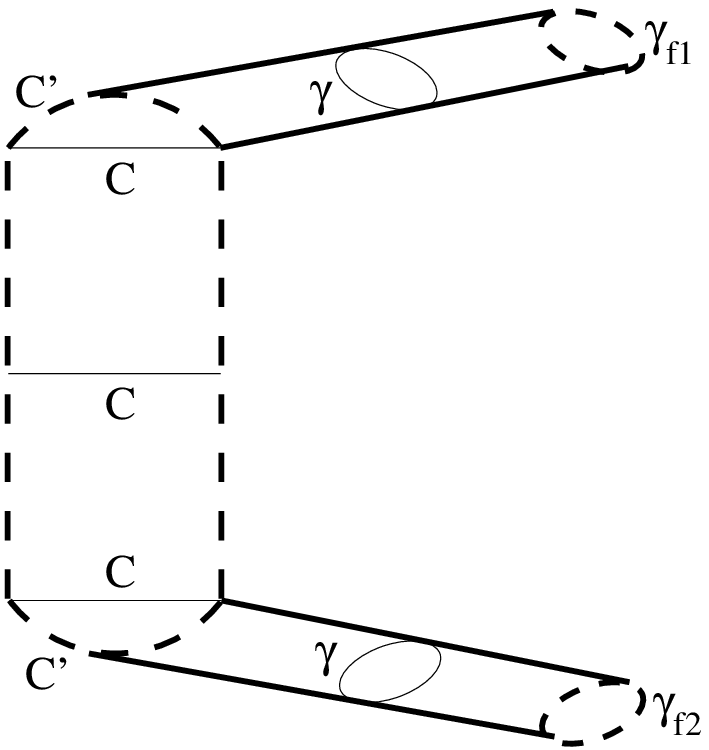}
\capl{f-mes}{An M2 brane representing a meson
(a ``blowup'' of figure \ref{f-scr}a).
\nl The dashed lines are boundaries (on $\Sg$); the thin lines are internal.}
\end{figure}
The physical and internal spaces are represented each by one direction.
We also suppress one of the dimensions of the M2 brane, so each point
represents a line (either open or closed).

We start with the left M2 brane (figure \ref{f-scr}a),
which represents a meson. A more detailed description of this M2 brane
is given in figure \ref{f-mes}.
The vertical line in figure \ref{f-scr}a,
extending in physical space, is the MQCD string;
each point on it is an open line (see figure \ref{f-mes}),
like $C$ in figure \ref{f-sus}.
Recall, however that it can be complemented by the curve $C'$ to a closed
curve%
\footnote{As remarked at the end of subsection \ref{sec-hom}, this
corresponds to a transition from an MQCD string to a fundamental string
and, therefore, should be highly suppressed dynamically.
However, the kinematical possibility is still important, since we
need to realize it only at {\em discrete points} -- the ends of the
MQCD string. The corresponding dynamical suppression is related to the
threshold for the production of a quark-anti-quark pair.}
$\gm_c$.
The horizontal lines, extending only in the internal space, are the quarks;
each point on them represents a closed loop $\gm$ (see figure \ref{f-mes});
these loops interpolate between
the loops $\gm_c$ and $\gm_f$, where $\gm_f$ is on a semi-infinite D4 brane.
In fact, the two quarks can have different $\gm_f$, being on different 
D4 branes (which means a different flavor, like in a $u\bar{d}$ meson).
The total M2 brane is, therefore, a surface with three boundaries:
$\gm_{f1}$, $\gm_{f2}$ and the third is composed of the two boundaries of
the MQCD string connected by two copies of $C'$.

The next M2 brane (figure \ref{f-scr}b)
describes an intermediate configuration. The additional
horizontal line is a cylinder in internal space which starts at $\gm_c$
(at some point in physical space along the MQCD string), passes through
some closed loop $\gm$ and goes back to $\gm_c$.
At the initial state, $\gm=\gm_c$; at an intermediate state, $\gm$ is
arbitrary (but homologic to $\gm_c$);
at the final state (figure \ref{f-scr}c), $\gm$ arrives to
$\gm_{f3}$, which is on a semi-infinite D4 brane and, therefore, can be
a boundary for the M2 brane. So at this stage, the two parts are free to
separate in space (as described in figure \ref{f-scr}d),
each representing a full meson.

So, the introduction of dynamical quarks, makes the MQCD string breakable.
This was shown both by topological considerations and
by an explicit construction.

\subsection{Including an Adjoint Field}
\secl{sec-adj}

So far we mostly considered a model obtained from the $\Nc=2$ supersymmetric
one, by decoupling the adjoint field $\Phi$ -- giving it an infinite mass.
However, the above analysis equally applies to a more general situation,
as we will now show.
We return to the $\Nc=2$ supersymmetric model and break the $\Nc=2$
supersymmetry by the following superpotential:
\beql{SuPot-Phi}
\Dl W=\tr f(\Phi) \hsc f(\Phi)=\sum_{k=2}^{N-1}\mu_k \Phi^k \hs.
\eeq
\internote{What happens when $k\ge N$?}
Such a perturbation has the same effect as a mass for the adjoint:
the coulomb branch is lifted and in the surviving vacua monopoles condense,
causing confinement.

The brane configuration corresponding to this model was found in \cite{BO9708}.
The superpotential (\ref{SuPot-Phi}) is realized by a ``curved'' left NS5
brane%
\footnote{In a certain singular limit this NS5 brane, curving back and forth
in the $w$ direction, can be seen as a set of parallel NS5 branes extending
in this direction.
This configuration was considered in \cite{EGK9702,EGKRS9704}.},
described asymptotically by the relation $w=f'(v)=\sum k\mu_kv^{k-1}$.
In semi-classical M theory there is, as before, a single M5 brane
$\RR^{3+1}\times\Sg$.
We are interested in curves $\Sg$ with vanishing
genus -- corresponding to vacua with no massless photons.
Imposing the appropriate asymptotic conditions, one obtains
\beql{Sg-SU2}
v=z+\frac{\lm}{z} \hsc w=[f'(v(z))]_> \hsc t=z^{N-F}\det(z+Z) \hs.
\eeq
Here $z$ is a global coordinate of the curve,
$Z$ is an $F\times F$ matrix related to the masses of the quarks
and $\lm^{N-F}=\Lm_2^{b_2}/\det Z$, where $\Lm_2^{b_2}$ is the
instanton factor of the $\Nc=2$ model ($b_2=2N-F$).
$[g(z)]_>$ means to expand the function $g(z)$ in powers of $z$
and to take only the {\em positive} powers. To relate this curve to the
previous one (eq. (\ref{Sg-SU})), one should choose $f(\Phi)=\half\mu\Phi^2$.
This implies $z=w/\mu$ and, therefore, leads to the curve
\beql{Sg-SU3}
\mu^N t=w^{N-F}\det(w+M) \hsc v=\frac{\zt}{w}+\frac{w}{\mu} \hs,
\eeq
where
\[ \zt:=\mu\lm=\left[\frac{\Lm^b}{\det M}\right]^\frac{1}{N-F} \hsc
   M:=\mu Z \hsc \Lm^b:=\mu^N\Lm_2^{b_2} \hsc
    \hs. \]
Taking the limit $\mu\goto\infty$ with $M,\Lm$ fixed
(and with $(\mu/\zt)^N t\goto t$),
one obtains the curve (\ref{Sg-SU}) with $m=-\zt M\inv$.

With this background, one can turn to the analysis of the MQCD string and
one finds that nothing essential is changed.
The space $\Mc$ is the same and the curve $\Sg$,
although more distorted in the $w$ direction, can still be identified with
a punctured plane -- this time the $z$ plane. Moreover, the behavior of $t$
is completely unaffected by the details of $\Dl W$ and, since this is
the only element relevant to the topological analysis, the results are the
same: for $F=0$ there is an MQCD string stabilized by a conserved $Z_N$
charge. The quark is realized in the same way (note that the semi-infinite
D4 brane realizing it is connected to the right NS5 brane, which is not 
affected by $\Dl W$), so the description of confinement and screening is also
the same. This is all in agreement with the expectations from field theory.

\newsection{The $\prod_\al SU(N_\al)$ Models}
\secl{sec-lSU}

One can generalize the models of the previous section by adding more NS5
branes%
\footnote{Such models were considered in
\cite{Witten9703,BH9704,BSTY9704,AOT9707,NOS9707,GP9708,ENR9801}.}.  
We consider a series of $l$ NS5 branes, each having fixed $x^6$ and
two dimensions extended in the $(v,w)$ direction.
There are also $N_\al$ D4 branes connecting the $\al$th and the $(\al+1)$th
NS5 brane,
$N_0$ semi-infinite D4 branes extending to the left of the first (left)
NS5 brane and $N_l$ semi infinite D4 branes extending to the right of the
last (right) NS5 brane.
When the NS5 branes are parallel, the low energy effective field theory on
the finite D4 branes is an $\Nc=2$ 4D gauge theory with a gauge group
$\prod_{\al=1}^l SU(N_\al)$.
For each $1<\al\le l$, there is a hypermultiplet in the
$(N_{\al-1},\wbar{N}_\al)$ (bi-fundamental) representation.
In addition, the semi-infinite D4
branes contribute each a hypermultiplet in the fundamental representation of
either $SU(N_1)$ (the left ones) or $SU(N_l)$ (the right ones).
Rotations of the NS5 branes correspond, as before, to masses for the scalars
in the adjoint representations and we will consider rotations that respect
$N=1$ supersymmetry.
In the semi-classical M-theoretical description, this is a single
M5 brane $\RR^{3+1}\times\Sg$.
For generic angles between the NS5 branes, one expects $\Sg$ to be
of genus 0, and  by imposing the appropriate asymptotic condition, one
obtains the following curve
\beql{Sg-lSU1}
v=\sum_{\al=1}^l \frac{c_\al}{z-z_\al} \hsc
w=\sum_{\al=1}^l \frac{s_\al}{z-z_\al} \hs,
\eeq
\beql{Sg-lSU2} t=\frac{\det(z-Z_R)}{\det(z-Z_L)}
     \prod_{\al=1}^l(z-z_\al)^{N_{\al-1}-N_\al}\hs.
\eeq
Here $z$ is a global coordinate of the curve, $c_\al,s_\al$ determine the
asymptotic position of the NS5 branes (which are related to the masses of the
adjoint scalars and the scales of the gauge factors), and $z_\al,Z_L,Z_R$
are related to the masses of the bi-fundamentals, left fundamentals and
right fundamentals, respectively.

\subsection{The MQCD string}

We start with the model with no fundamental matter ($N_0=N_l=0$). For generic
adjoint masses, the model is expected to be in the confining phase.
The elements of $SU(N_{\al-1})\otimes SU(N_\al)$ acting trivially on the
bi-fundamental representation are of the form
$(\om I_{N_{\al-1}},\om I_{N_\al})$,
so the kernel of this representation is $\ZZ_N$, where
$N=\gcd(N_{\al-1},N_\al)$
is the greatest common divisor of $N_{\al-1}$ and $N_\al$.
Therefore, for the above
model, the kernel is $\ZZ_N$ with $N=\gcd(N_1,\ldots,N_{l-1})$,
embedded diagonally in all the group factors.
When $N>1$, one expects the existence of strings (flux tubes) with a conserved
$\ZZ_N$ charge, representing a confining force,
while when $N=1$, there should be no conserved charge,
which means that the confining force is screened. 

As before, we will identify the $\ZZ_N$ group with the relative homology group
$H_1(\Mc/\Sg)$. The manifold $\Mc$ is the same, leading to $H_1(\Mc)=\ZZ$.
To find $H_1(\Mc/\Sg)$ using eq. (\ref{HR}), we look for classes in 
$H_1(\Mc)$ that can be represented by loops in $\Sg$ (and are, therefore,
trivial in $H_1(\Mc/\Sg)$).
The surface $\Sg$, as described in eqs. (\ref{Sg-lSU1},\ref{Sg-lSU2})
(with $N_0=N_l=0$),
can be identified with the $z$ plane, complemented by $z=\infty$ (which is
a finite point on $\Sg$!) and punctured at the points $z_\al$,
where at least one of the space-time coordinates diverges.
Loops in $\Sg$ are classified by their
winding number around these punctures. From eq. (\ref{Sg-lSU2})
we learn that a loop around $z_\al$ winds the compact direction
$N_{\al-1}-N_{\al}$ times. All these loops together generate in $H_1(\Mc)$
the subgroup $N\ZZ$, where $N=\gcd\{N_{\al-1}-N_{\al}\}=\gcd\{N_\al\}$
and using eq. (\ref{HR}), we obtain
\[ H_1(\Mc/\Sg)=\ZZ_N \hsc N=\gcd\{N_\al\} \hs, \]
which is the desired result.

\begin{figure}
\pct{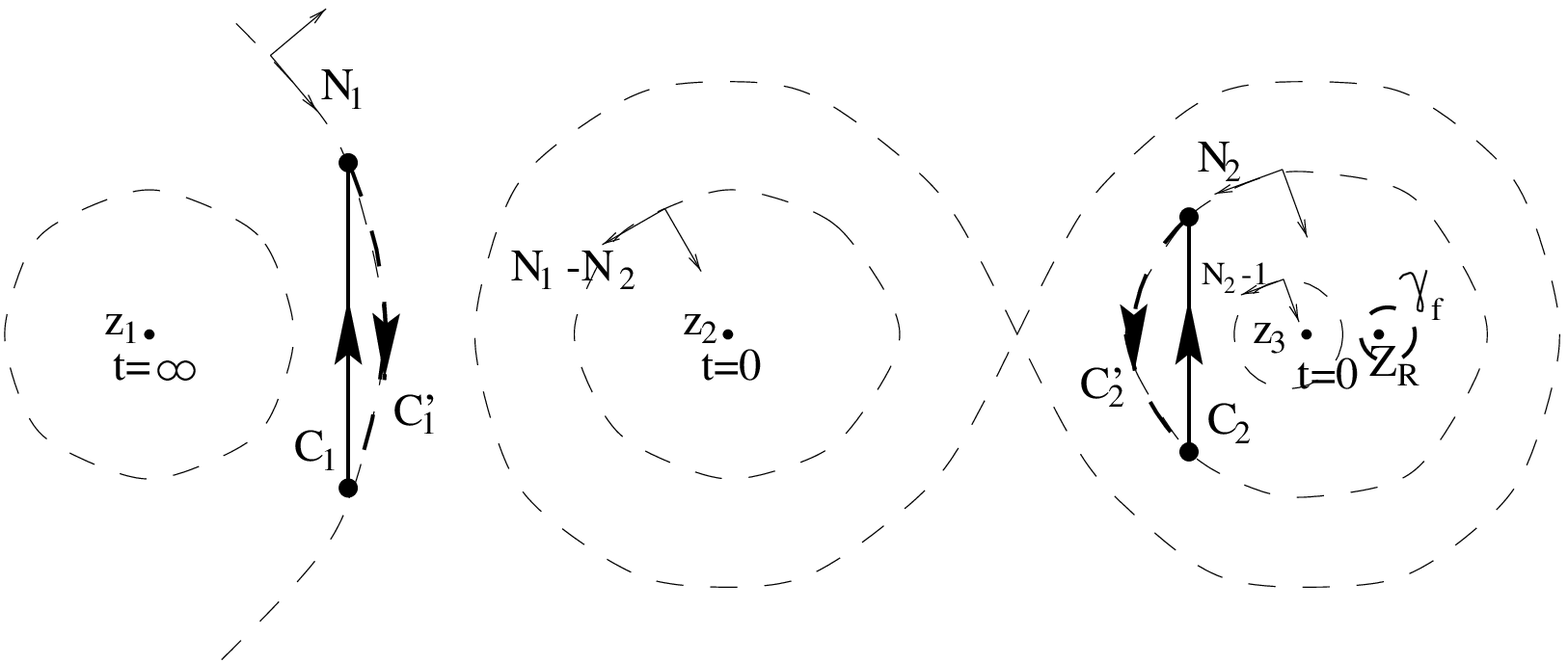}
\capl{f-lsu}{The curve $\Sg$ for $SU(N_1)\otimes SU(N_2)$ with $N_1>N_2$}
\end{figure}
Figure \ref{f-lsu} describes $\Sg$ for $SU(N_1)\otimes SU(N_2)$ gauge group.
It is the $z$ plane, with which $\Sg$ can be identified.
As in figure \ref{f-sus}, we concentrate on the relevant aspect, which is
the $t$ behavior and, in particular, its argument, which parameterizes the
winding of $\Sg$ around the compact dimension. The points $z_\al$ are the
asymptotic locations of the three NS5 branes. The dashed lines are lines of
constant $|t|$ (constant $x^6$).
Each is a closed curve (possibly passing through $z=\infty$).
Near some of these curves, there are arrows that indicate the
positive direction of $x^6$ and $x^{10}$ and a number $k$ which is the winding
number of this circle around $x^{10}$. There are $k$ points on this circle with
the same $x^{10}$ and, therefore, the same $t$.
The MQCD string $C$ may be chosen to connect two such points. For definiteness,
one can choose constant $t$ along it and interpolate linearly between the end
values of $v$ and $w$. Two such strings are described: $C_1$ between the
left D4 branes and $C_2$ between the right ones.
As for the $SU(N)$ gauge group, each of these strings can be complemented by a
line $C'$ on $\Sg$ to a
closed curve $\gm_c$ winding once around $x^{10}$.
The point $z=Z_R$ is the asymptotic location of a semi-infinite D4
brane connected to the right of the right NS5 brane and representing a quark
in the fundamental representation of $SU(N_2)$. A curve $\gm_f$ around this
point winds once around $x^{10}$.

Using this figure, we can demonstrate confinement and screening in this model,
along the lines of the previous section. An M2 brane describing a quark should
be a cylinder with a curve like $\gm_f$ as one of its two boundaries.
The other boundary $\gm_c$ should wind around $x^{10}$ once.
When $N_1$ and $N_2$ are relatively prime, one can construct $\gm_c$
by combining loops around $Z_1$, $Z_2$ and $Z_3$ (for example, when
$N_1-N_2=1$, one can take for $\gm_c$ a loop around $Z_2$).
So in this case, there are isolated quarks, which means that
the confining force is completely screened (loops around $Z_1$, $Z_2$ and
$Z_3$ correspond to screening by an adjoint of $SU(N_1)$, bi-fundamental and
adjoint of $SU(N_2)$ respectively).

When $N=\gcd(N_1,N_2)>1$, there is no curve on $\Sg$ that winds once around
$\gm_c$ (except at the region near $Z_R$, which is identified as a
{\em flavor} D4 brane and not a {\em color} one). A possible boundary is
$\gm_c=C+C'$, which corresponds to a quark attached to an MQCD string. So
we conclude that in this case quarks can not exist in isolation and they are
bound to ends of MQCD strings.
We chose $Z_R$ to be close to $z_3$, which means that
it is heavy and, therefore, decouples at low enough energies
(an infinite mass for this quark would correspond to $Z_R\goto z_3$).
It is manifested geometrically by the fact that away from $Z_R$ and $z_3$,
the curve is as for the model without fundamentals and an MQCD string is
stable there -- quarks are confined.
At higher energies, the region of $Z_R$ and $z_3$ becomes accessible to the
MQCD string and it can break.
This is demonstrated by figure \ref{f-scr} also for the present case,
where this time $\gm_f$ and $\gm_c$ are as in figure \ref{f-lsu}.
To summarize, also for this model, we obtained results in agreement with the
field theoretical expectations.

\newsection{Models with an O4 Orientifold}
\secl{sec-O4}

We now move to the analysis of confinement and screening in models with
$Sp(N)$ or $SO(N)$ gauge groups.
These models can be realized in type IIA string theory \cite{EJS9703,EGKRS9704}
by performing an
orientifold projection, with a 4D orientifold plane (O4 plane)
parallel to the D4 branes. 
The brane configuration is almost as in figure \ref{f-cfg}, the only
difference being that it is restricted to be symmetric around the O4 plane,
which is the $x^6$ axis (multiplied, as everything else, by $\RR^{3+1}$).
This means that the D4 branes can move off the $x^6$ axis only in symmetric
pairs. There are two possible types of projections, one leading to $Sp(N)$
gauge theory ($N$ even) with fundamental matter fields and the other,
to $SO(N)$ gauge theory with matter in the vector representation.
In both cases $F$ is even and there are $\half F$ $\Nc=2$ hypermultiplets.
The realization of fields and the identification of parameters and moduli is
as in the $SU(N)$ model, described in subsection \ref{SU-brane}.
So is the supersymmetry. 

These configurations also have a (semi-classical) M-theoretical description,
as a single smooth M5 brane \cite{LLL9705,BSTY9705}%
\footnote{M5 branes corresponding to models with O4 orientifolds were
constructed and analyzed also in \cite{CS9708}-\cite{Hori9805}.}.
As the type IIA
configuration, the M5 brane is also symmetric around the orientifold plane
(modulo a shift in the compact dimension, which is invisible in the type IIA
description). In addition, it was found that the O4 plane
contributes additional tubes to the M5 brane, as would do 
D4 branes with the same RR charge. The charge of an O4 plane depends on
the projection type. For an $Sp(N)$ projection it is $+1$ (in units of D4
brane charge) and for the $SO(N)$ projection it is $-1$. When an orientifold
plane crosses a NS5 brane, the sign of its charge changes
\cite{EJS9703}, so the NS5 brane feels a net charge of two units.
This implies that in the $Sp(N)$ model there are two extra tubes
between the NS5 branes and in the $SO(N)$ model there are two extra tubes
in the semi-infinite regions. These tubes cannot move off the $x^6$ axis.

When the NS5 branes are not parallel, all the finite tubes merge to a
single tube, as in the $SU(N)$ model;
for orthogonal NS5 branes, corresponding to an
$\Nc=1$ model without an adjoint, $\Sg$ assumes the following form
\cite{CS9708,AOT9708,AOT9709}
\beql{Sg-SOp}
t=v^{-k}\det(m-v)/\det m \hsc vw=\zt:=[\Lm^{3k-2F}\det m]^{\frac{1}{k}} \hs,
\eeq
where $k=N+2$ for $Sp(N)$ and $k=N-2$ for $SO(N)$.
The corresponding  M5 brane is invariant under the transformation
\beql{reflect-M} 
v\goto-v \hsc w\goto-w \hsc x^7\goto-x^7 \hsc t\goto(-1)^N t\hs. \eeq

These models are also expected to be confining, so we would like to find the
corresponding MQCD string, following the considerations in subsection
\ref{sec-hom}. Since this should be an M2 brane, we have to determine first
what is the effect of an orientifold projection on such branes.
This will be considered in the next subsection.

\subsection{The M2 brane configurations}
\secl{s-M2}

In the M theoretical description of the O4 orientifold, the transformation
$\Om$ which is gauged involves not only the space reflection 
(\ref{reflect-M}), but also an orientation reversal of the M2 brane.
\internote{This implicitly assumes orientability.
\nl Note that we consider surfaces in $\what{\Mc}$ which is the covering space
(see below) and, presumably, the M2 branes in $\what{\Mc}$ indeed are
orientable, as are the fundamental strings before the $\Om$ projection
(these are type II strings!)
\nl Anyway, for the MQCD string, we have $\RR\times\what{C}$. with $\what{C}$
an $\Om$-symmetric curve and a curve (with  no singular points)
cannot be unorientable!}
One way to see this is from 11D supergravity --
the low energy approximation of M theory. To gauge $\Om$, it has to
be a symmetry of the corresponding action. This action includes a term
\beql{SUGRA} \int A\wedge dA\wedge dA \hsc \eeq
where $A$ is a 3-form field. The transformation (\ref{reflect-M}) reverses the
orientation of the 11D space-time (inverts an odd number of coordinates), so
to keep the term (\ref{SUGRA}) $\Om$-invariant, $A$ should change sign
(as already observed in \cite{HW9510,DM9512,Witten9512}).
The M2 brane couples to $A$ through a term $\int_{\rm M2}A$, and to keep this
term invariant too, the orientation of the M2 brane should be inverted.
The same conclusion is obtained by considering the brane realization of
$N=2$ SUSY quantum gauge theory (corresponding to parallel NS5 branes).
The central charge of a BPS state is given by $\int_\gm\lm$,
where $\lm=vdt/t$ is the Seiberg-Witten differential and $\gm$ is a
homology cycle in $\Sg$. In M theory, a BPS state is realized by an M2
brane $S$ with boundaries on the M5 brane (\ie\ on $\Sg$), the homology cycle
$\gm$ is the boundary of $S$ and the integral gives
the (signed) surface of this brane.
\internote{More details:
\nl Area$\ge|Z|$, $Z=\int_S\Om_S=\oint_{\pt S}\lm$, $\Om_S=d\lm$.
\nl $\Om_S$ is the pullback to $S$ of the covariantly constant holomorphic
2-form $\Om$.
\nl In a space $F(y,z,v)=0$, $\Om=dv\wedge dy/(\pt F/\pt z)$.
\nl For $F=yz-P(v)$, $\Om=\frac{dv\wedge dy}{y}=d\left(\frac{vdy}{y}\right)$.} 
In an orientifold model,
the $\Om$-mirror $\wbar{S}$ of $S$ represents the same state and thus should
lead to the same central charge.
Now one observes that $\lm$ is odd under the transformation
(\ref{reflect-M}), while $\Sg$ is symmetric%
\footnote{The specific form of $\Sg$ for the $\Nc=2$ models \cite{LLL9705}
is not needed here.},
so the integral $\int_\gm\lm$ changes sign when $\gm$ is transformed by
(\ref{reflect-M}) and, to obtain the same central charge, the transformation
(\ref{reflect-M}) must be accompanied by a flip of orientation.
Finally, the flip in orientation is also consistent with the limit of
(perturbative) type IIA string theory.
There, by definition, $\Om$ flips the orientation of the
fundamental string. In M theory, a fundamental string corresponds to a
tube-like M2 brane wrapped around the compact dimension,
parameterized by $x^{10}$. The transformation (\ref{reflect-M}) does not
change the $x^{10}$ direction, so the direction of the string is
correlated with the orientation of the corresponding (tube-like) M2 brane.
\internote{This arguments applies directly only to `stringy' M2 branes
(those becoming fundamental strings in the $R\goto0$ limit) and
their continuous deformations.
\nlb It turns out, however, that all the M2 branes we consider are of this
type.
\nlb The other two arguments apply to all M2 branes, including those that
become D2 branes in the $R\goto0$ limit (\ie, do not wrap around $x^{10}$).
M2 branes of this kind represent magnetically charged BPS states.}

For the analysis of the MQCD string, we need a characterization of the space
of M2 brane configurations in a model with an O4 orientifold%
\footnote{This is the space on which a wave functional of the brane would
be defined. Alternatively, in a path-integral formulation, this would define
the domain of integration.}.
If this was an ``orbifold model'', obtained by gauging the reflection
(\ref{reflect-M}), the answer would be clear: one would consider general M2
brane configurations in the {\em quotient space} $\Mc\times\RR^3$, where
\beql{Mc} \Mc=\what{\Mc}/\ZZ_2 \hsc \what{\Mc}=\RR^6\times S^1 \eeq
and $\ZZ_2$ is the reflection (\ref{reflect-M}).
In the present case, as explained above, the gauge transformation $\Om$
involves also an orientation reversal of the M2 brane,
leading to an undetermined orientation in the quotient space, so to keep the
information about the orientation, one has to consider also the 
{\em covering space} $\what{\Mc}\times\RR^{3+1}$.
\internote{Alternatively, one could consider surfaces directly in the
quotient space $\Mc$.
\nl [Witten9706]: One should consider homology groups with {\em twisted}
coefficients (\ie, transforming non-trivially ($n\goto-n$?) under $\Om$).}
In the covering space,
one should consider general configurations, with the understanding that
configurations related by $\Om$ should be identified.
One should observe that there are closed configurations in the quotient space
$\Mc\times\RR^{3+1}$ that are represented in the covering space
$\what{\Mc}\times\RR^{3+1}$ by configurations that are {\em not} closed,
\ie, ``jump'' between points identified by $\Om$
(the ``twisted'' sector; leading to unoriented surfaces in
$\Mc\times\RR^{3+1}$) so, in general, one should consider also such
configurations.

We now specialize to a configuration representing an MQCD string, looking for
the conserved charge realizing the flux (recall the discussion in subsection
\ref{sec-hom}). These are M2 branes ending on the M5 brane but, as in the
previous models, one can start by classifying M2 branes in the absence of the
M5 brane.
To find a candidate for a conserved charge, we
consider first the case of a closed M2 brane in the covering space
(corresponding to an orientable M2 brane in the quotient space). 
The internal space is $\what{\Mc}=\RR^6\times S^1$, the same as in the
$SU(N)$ case, so these M2 branes seem to be characterized, as before,
by the winding number of the cross-section $\tilde{C}$ around the compact
dimension, leading to $H_1(\what{\Mc})=\ZZ$.
However, here the situation is different.
This is because we look for a physical property, which must be
$\Om$-invariant (since this is a gauge symmetry), while the winding number is
not. To analyze this, it is worthwhile to consider the quotient space.
Denoting by $C$ the curve in
$\Mc$ that is the image of $\tilde{C}$, the winding number is 
\beql{wind} \frac{1}{2\pi R}\int_{\tilde{C}}dx^{10}=
   \frac{1}{2\pi R}\int_{C}dx^{10} \hs, \eeq
where the integration over $C$ is performed using the orientation induced on
it by that of $\tilde{C}$. Different curves $\tilde{C}$ in $\what{\Mc}$
which are related by $\Om$, map to the same $C$ in $\Mc$, but induce
on it a different (continuous) orientation, leading to a different winding
number.
For example, replacing $\tilde{C}$ by its $\Om$ image leads to the opposite
orientation on $C$. Since $dx^{10}$ is $\Om$ invariant, the orientation flip
implies an opposite winding number.
This means that only the winding number modulo 2
{\em can} be gauge invariant and we now show that {\em it is} indeed gauge
invariant. To show this, we should compare two different orientations. 
In general, the orientations can be different only on a part of $C$.
This part is, however, necessarily closed (since both
orientations are continuous), so it contributes an integer to the winding
number. Therefore, when its orientation is flipped, the change in the winding
number is even, so the winding number modulo 2 is invariant.
The above argument also demonstrates that the winding number modulo 2
can be calculated directly in the quotient space $\Mc$, by choosing an
{\em arbitrary} continuous orientation on $C$.

We return now to the general M2 brane, which can be also unorientable. The
lack of orientability is a global property of the surface and will not be
reflected in the classification of cross-sections, so these will still be
classified by the winding number modulo 2. 
The results obtained so far can be summarized as follows:
\begin{itemize}
\item
The gauge invariant property that can be used to classify the cross-sections
of an MQCD strings is a $\ZZ_2$ charge:
the winding number modulo 2 of the cross-section around the compact
dimension. 
\item
This property can be calculated directly in the quotient space $\Mc$,
using eq. (\ref{wind}) with an arbitrary (continuous) orientation on the
cross-section $C$.
\end{itemize}
\internote{Mathematically, this means that the physically relevant
homology group is that with $\ZZ_2$-valued coefficients (rather then integer
coefficients).
\nlb Actually, the homology group of unoriented spaces is necessarily with
$\ZZ_2$ valued coefficients, since the sign, which is related to the
orientation, is undetermined.
\nlb This should be somehow related to the fact that a boundary of an
unoriented surface can have a non vanishing (but even) winding number around
a circle.}

Next we analyze the dependence of the $\ZZ_2$ charge on the cross-section.
For this, let $C_1$ and $C_2$ be two cross-sections (in the quotient space)
and $S$ the part of the M2 brane between them (so $C_2-C_1$ is the boundary
of $S$). If $S$ is orientable, one can choose a global orientation and this
leads, as in the $SU(N)$ case, to the same winding number for any two
cross-sections, so the $\ZZ_2$ charge is trivially independent of the
cross-section. 
To check the unorientable case, let us represent it as an orientable surface
with insertions of cross-cups.
\internote{An unorientable surface can always be represented (embedded in
$\RR^3$?) as an orientable one with a single cross-cup. What happens to the
winding number?}
Assuming even $N$ in the transformation (\ref{reflect-M}) (odd $N$ will be
considered in subsection \ref{Odd-N}), $x^{10}$ is a gauge invariant
function on $S$. A difference in the winding numbers of $C_1$ and $C_2$ can
be traced to a non-trivial winding for a loop around a cross-cup.
However, since opposite points on a cross-cup are identified and,
therefore have the same $x^{10}$ value, the winding number for a loop around
the cross-cup is always even, so $C_1$ and $C_2$ have the same $\ZZ_2$ charge
also in this case. We, therefore, have a well defined, gauge invariant
$\ZZ_2$ charge for the MQCD string. To proceed, we consider each gauge
group separately.

\subsection{The $Sp(2r)$ models} 
\secl{sec-Sp}

As for $SU(N)$, we first consider the model without hypermultiplets ($F=0$).
All the fields are in the adjoint representation, which is invariant under
the center of the group.
The center of  $Sp(2r)$ is $\ZZ_2$, so we expect an MQCD string with
such a charge.
This is indeed what we find. As in the previous models, one can consider
only closed cross-sections, by supplementing open ones with curves in $\Sg$.
These were analyzed in the previous subsection and we obtained a $\ZZ_2$ charge
that, in the absence of an M5 brane, was conserved. The M5 brane can lead to
non-conservation by adding a loop in $\Sg$ to a cross-section $C$.
The curve $\Sg$ is given (in the covering space) by eq. (\ref{Sg-SOp})
(with $F=0$), leading to identification modulo $k$ in the winding number.
$k=2r+2$ is even so the $\ZZ_2$ charge is still conserved.
this $\ZZ_2$ charge can, therefore, be identified with the center of the
$Sp(N)$ gauge group.

Figure \ref{f-sps} describes two examples of an MQCD string (or, more
precisely, their cross-sections in the covering space), with a unit charge.
\begin{figure}
\pct{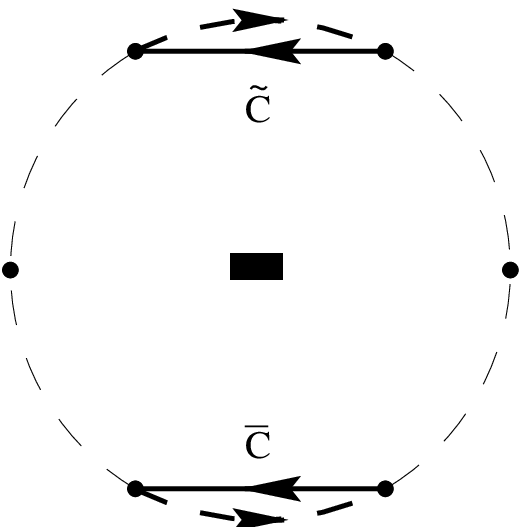}
\capl{f-sps}{MQCD strings for the $Sp(N)$ model}
\end{figure}
This looks exactly as for $SU(N)$ (compare to figure \ref{f-sus}), but in that
case $\tilde{C}$ and $\wbar{C}$ are distinct and carry different (opposite)
$\ZZ_N$ charge (winding number), while in the present case, they are
$\Om$-equivalent, carrying {\em the same} ($\ZZ_2$) charge.
It is easy to see how two such strings can annihilate each other:
$\wbar{C}$ can be rotated until it coincides with $\tilde{C}$ and
they can join to a loop with a vanishing winding number, which can contract
to a point.

Confinement and screening are analyzed in the same way as in $SU(N)$, by
introducing semi-infinite D4 branes, representing heavy quarks.
Also here M2 branes representing isolated quarks look as in figure
\ref{f-qrk} and are possible only when the NS5 branes are parallel.
When the NS5 branes are rotated, the color D4 branes join to a single tube
which winds around $x^{10}$ more then once, so a quark tube cannot end on
it. Instead, it can connect to an MQCD string, whose cross section in the
internal space is described in figure \ref{f-sps}. So quarks are
connected in pairs by a flux tube, forming singlets.
Note that unlike $SU(N)$,
in $Sp(N)$ there is no distinction between the fundamental
and anti-fundamental representations and any two quarks can combine to form a
singlet (being the analog of both mesons and baryons in $SU(N)$).
This is reflected correctly in the M theory description.
In $SU(N)$, the two representations were distinguished by the orientation of
the corresponding M2 branes and an MQCD string could connect only quark M2
branes of opposite orientation. Here, the two orientations are related by
$\Om$ so the MQCD string can connect any two quark M2 branes.

\internote{In the symmetric approach:
\nlb
Here the M2 branes come (in the covering space) in pairs with opposite
orientation and an MQCD string can connect any two such pairs.
\nlb
Alternatively, in the quotient space, the M2 brane has no orientation.}
As in previous models, at low energies, compared to the mass of the quarks,
the MQCD string is stable and there is confinement.
At higher energies, pairs of quarks can be produced dynamically, screening
the confining force. This process is described by figure \ref{f-scr} also for
the present case.

To summarize, we have reproduced all the results described in section 2,
also for the symplectic groups. We now turn to the orthogonal groups.

\subsection{ The $SO(2r)$ models ($r>1$)} 

The group $SO(N)$ is not simply connected: it is covered twice by its universal
covering group $Spin(N)$, so $SO(N)=Spin(N)/\ZZ_2$.
$Spin(2r)$ has a center of four elements - either $\ZZ_4$
(for odd $r$) or $\ZZ_2\times\ZZ_2$ (for even $r$), so this should be the
charge of a general string in the model without hypermultiplets ($F=0$).
The search for the MQCD string is identical to the one for $Sp(2r)$, with
the same results: MQCD strings with a $\ZZ_2$ charge. This means that we do not
obtain the full spectrum of possible MQCD strings, but actually this could be
expected. A string (flux tube) with a unit charge would have to be connected to
$SO(N)$ spinors. These are not realized in the present brane configuration,
so it is reasonable that the corresponding strings are not realized either.

We conclude that we have found an MQCD string carrying the charge of the
{\em vectorial} representation of $SO(2r)$.
This is verified when we add semi-infinite D4 branes, which
realize quarks in this representation:
the M2 branes realizing the quark states are bound in pairs by such a string,
forming a meson, as in the $Sp(2r)$ model.
The demonstration of confinement and screening is also identical in the
two models. Note, however, that not all charges are screened:
the dynamical matter can screen (for $F>0$) any external probe
which is in a representation of $SO(2r)$ (rather then $Spin(2r)$). When
spinorial external probes are available, they should feel a confining force,
but we do not have a description of those in the present formulation,
so this cannot be checked.
\internote{Representation of baryons?}

\subsection{The $SO(2r+1)$ models} 
\secl{Odd-N}

The center of $Spin(2r+1)$ is $\ZZ_2$ and the vectorial representation is 
uncharged under it (being a representation of $SO(2r+1)=Spin(2r+1)/\ZZ_2$).
Therefore, in view of the discussion of the $SO(2r)$ models, we expect here
no MQCD string. To check this, we return to the discussion at the end of
subsection \ref{s-M2}. The present case corresponds to odd $N$ in the
transformation (\ref{reflect-M}), meaning that an orientation flip in $S$ is
accompanied by a jump of $\pi R$ in $x^{10}$. A loop around a cross-cup will,
therefore, have an {\em odd} winding number and this implies that there will
always be cross-sections of the MQCD string with an even, \ie\ trivial,
winding number.
Thus, in the present case, although the $\ZZ_2$ charge of a specific
cross-section of the string is well defined, it does not protect the
string from breaking, since it is not the same for all cross-sections.
This means that there is no stable MQCD string, in agreement with the field
theoretical expectations.
Note that we obtained this result even before considering the M5 brane!
For completeness, let us nevertheless consider the influence of $\Sg$.
It leads, as before, to identification
modulo $k$ in the winding number and, since this time $k=2r-1$ is odd, the
$\ZZ_2$ charge of a cross-section is no longer conserved.
\internote{In the symmetric approach:
\nl
We have two sources for non-conservation: touching $\Sg$ changes the charge
by multiples of $k=2r-1$, while $\tilde{C}$ and $\wbar{C}$ can exchange
charge by multiples of 2.
Unlike the previous cases, now $k$ is odd, so by combining
these two processes, we see that nothing is conserved.}
As for $SO(2r)$, the absence on a stable MQCD string
means that matter in representations of $SO(2r+1)$ (and, in particular, the
vectorial one) is screened. Spinorial charge is expected to be confined.

\newsection{Summary}
\secl{sec-disc}

In this work we used the realization of supersymmetric gauge theories in M
theory to investigate confinement and screening. We considered the following
($\Nc=1$ supersymmetric) models:
\begin{itemize}
\item 
$\Nc=2$ SUSY $SU(N)$ gauge theory with $F$ fundamental flavors, perturbed by
a polynomial superpotential for the adjoint hypermultiplet
($\Nc=1$ SQCD can be seen as a special case of this family, for which the
adjoint hypermultiplet is infinitely heavy);
\item
$\Nc=2$ SUSY $\prod_\al SU(N_\al)$ gauge theory with matter in fundamental and
bi-fundamental representations, perturbed by masses for the adjoint
hypermultiplets;
\item
$Sp(N)$ gauge theory with $F$ fundamental hypermultiplets;
\item
$SO(N)$ gauge theory with $F$ vectorial hypermultiplets.
\end{itemize}
For all the above models, we found results in agreement with the field
theoretical expectations.
Starting without fundamental matter (vectorial, for $SO(N)$), the MQCD string
(\ie, a candidate for the field theoretical electric flux tube) was identified
as an M2 brane ending on the M5 brane (following \cite{Witten9706}).
This string was shown to carry a topologically-conserved charge under a group
isomorphic to the group $H$ of gauge transformations that act trivially on the
fields in the field theory.
This is in accord with the identification of this charge as the gauge charge
that is {\em not screened} by the dynamical matter and, therefore, confined.
We then introduced heavy quarks carrying unscreened charge and showed that
they cannot exist in isolation, but can be connected to an MQCD string, forming
neutral mesons. This is a demonstration of confinement.
Finally, we considered higher
energy scales, at which these quarks become dynamical and
found that this destabilizes the MQCD string, demonstrating the screening of
the charge that these quarks carry.

In the $SO(N)$ models, we did not obtain strings carrying spinorial charge: for
example, for $SO(2r)$, the MQCD string carries a $\ZZ_2$ charge (which was
shown to correspond to the vector representation), while the
field theory expectation is for a charge group $H$ with 4 elements. 
The fact that the charge group was at most $\ZZ_2$, followed from the topology
of the internal space $\Mc$ (as implied by the O4 projection), 
and was independent of the M5 brane. This can serve as guidance in the search
for a realization of models with spinors: the presence of heavy%
\footnote{In a configuration in which the vectors, for example, are restricted
to be massless, one expects a $\ZZ_2$ charge, since the vector screens the
rest.}
spinors and vectors would imply the existence of a $\ZZ_4$ or
$\ZZ_2\otimes\ZZ_2$ charge group and to obtain such a charge, a different
background is needed (\eg, larger orientifold group).
A different configuration of NS5 branes and D4 branes would correspond to a
different M5 brane in {\em the same background}, so it will not lead to the
desired result. 

The above considerations demonstrate that the properties
of the MQCD string can be used in the identification of the field theoretical
model realized by an M5 configuration, or serve as a consistency check. It is,
therefore, worthwhile to perform this analysis in other such realizations.

%%%%%%%%%%%%%%%%%%%%%%%%%%%%%%%%%%%%%%%%%%%%%%%%%%%%%%%%%%%%%%%%
\vspace{1cm}
\noindent{\bf Acknowledgment:}
We are grateful to A. Giveon, A. Hanany, M. Henningson, K. Landsteiner,
R. Livne, E. Lopez, Y. Oz and A. Zaffaroni for helpful discussions.
O.P. is grateful to the Theory Division at CERN, where part of this work
was performed, for hospitality.
This work is supported in part by BSF -- American-Israel Bi-National
Science Foundation, and by the Israel Science Foundation founded by the
Israel Academy of Sciences and Humanities -- Centers of Excellence Program.

%%%%%%%%%%%%%%%%%%%%%%%%%%%%%%%%%%%%%%%%%%%%%%%%%%%%%%%%%%%%%%%%
%% APPENDICES

\appendix
\renewcommand{\newsection}[1]{
 \vspace{10mm} \pagebreak[3]
 \refstepcounter{section}
 \setcounter{equation}{0}
 \message{(Appendix \thesection. #1)}
 \addcontentsline{toc}{section}{
  Appendix \protect\numberline{\Alph{section}}{#1}}
 \begin{flushleft}
  {\large\bf Appendix \thesection. \hspace{5mm} #1}
 \end{flushleft}
 \nopagebreak}

%%%%%%%%%%%%%%%%%%%%%%%%%%%%%%%%%%%%%%%%%%%%%%%%%%%%%%%%%%%%%%%%
%% TEXT OF APPENDICES
%%  use \newsection instead of \section

\ifinter
\internote{Appendices are unchanged since 10/5 and are omitted.}
\input qcdsup
\fi
%%%%%%%%%%%%%%%%%%%%%%%%%%%%%%%%%%%%%%%%%%%%%%%%%%%%%%%%%%%%%%%%
%% BIBLIOGRAPHY

\end{document}